\documentclass{ws-procs975x65}

\begin{document}

\wstoc{Evolving relativistic fluid spacetimes spectrally}{M. Duez}

\title{EVOLVING RELATIVISTIC FLUID SPACETIMES \\
  USING PSEUDOSPECTRAL METHODS AND FINITE DIFFERENCING}

\author{MATTHEW D. DUEZ, LAWRENCE E. KIDDER, SAUL A. TEUKOLSKY}

\address{Department of Astronomy, Cornell University, Ithaca, NY 14850 USA\\
\email{mduez@astro.cornell.edu}}

\begin{abstract}
We present a new code for solving the coupled Einstein-hydrodynamics equations
to evolve relativistic, self-gravitating fluids.  The Einstein field equations
are solved on one grid using pseudospectral methods, while the fluids are
evolved on another grid by finite differencing.  We discuss implementation
details, such as the communication between the grids and the treatment of
stellar surfaces, and present code tests.
\end{abstract}

\bodymatter

\section{Introduction}\label{intro}

Numerical relativity has the potential to make indispensable contributions
to our understanding of hydrodynamic compact object phenomena, such as
neutron star-neutron star (NSNS) binary
merger, black hole-neutron star (BHNS) binary merger, and stellar core
collapse.  In such systems, both the
spacetime metric and the fluid are dynamical, and they are strongly coupled. 

The most common approach to numerically solving the coupled
Einstein-hydrodynamics equations is by finite differencing. 
Finite difference (FD) techniques have been successfully used to
simulate NSNS binaries, stellar collapse, and other interesting
phenomena.   FD algorithms
usually converge to the exact solution as some power of the grid spacing. 
In addition, techniques have been developed which can evolve fluids with
discontinuities stably and accurately.  Unfortunately, FD codes usually
require very large grids in order to obtain accurate results.

Einstein's equations can also be evolved using pseudospectral (PS) methods. 
For smooth functions, PS methods converge {\it exponentially}
to the exact solution as the number of
collocation points is increased.  This allows PS methods to
get accurate results with much smaller grids than those used by FD codes. 
A PS code for solving the Einstein equations has been developed by the
Cornell-Caltech relativity group~\cite{spec1,spec2} and
successfully used to carry out binary black hole inspiral
simulations~\cite{BBH} which are both the most accurate and
computationally cheapest of their kind.

There is a difficulty in evolving non-vacuum spactimes spectrally, however. 
Because of the possibility of stellar surfaces
and shocks, the evolved variables are not always
smooth at all derivatives.  In these cases, spectral representations
display Gibbs oscillations
near the discontinuity which converge away only like a power of the
number of collocation points, the order of convergence given by the order of
the discontinuity.  In some cases, the problem can be avoided by
placing domain boundaries at discontinuities, but 
this is not practical for complicated shocks or
strongly deformed stellar surfaces.

Another possibility would be a mixed approach: to evolve 
the metric fields, which are much smoother, using PS methods
and evolve the
hydrodynamic fields using shock-capturing FD methods.  This would seem to
utilize the strongest features of each method.   This approach has been
used successfully in a conformal gravity code to perform stellar collapse
computations~\cite{MofM}.  Here we extend this approach to full GR. 

\section{Numerical Algorithm}\label{code}

We integrate the hydrodynamic equations in conservative form. 
using piecewise parabolic reconstruction~\cite{PPM}
together with a high-resolution central scheme~\cite{kt00}.  We use uniform
grids in three or two dimensions
(the latter for axisymmetric systems).  The vacuum outside the stars is
handled by introducing a tenuous ``atmosphere'', together with a density floor
and an internal energy ceiling, in these regions. 
The Einstein equations are evolved in the generalized harmonic
system~\cite{spec2}.  We find that filtering the metric variables is
sufficient to stabilize the PS code in the presence of discontinuities. 
The interpolation from the spectral to the fluid grid
would be very expensive if done directly, but we make the process much
quicker using a technique introduced by Boyd~\cite{b92}.  Both codes
use dual coordinate frames, which allow the grid to dynamically adjust
to the motion of the system~\cite{BBH}.

Our finite difference code currently has no adaptive griding capability,
but our two-grid approach gives us a few similar advantages. 
The finite difference grid need only cover the region containing the matter--
a huge savings in many binary applications.  The dual coordinate frame system
allows the finite difference grid to move with the stars.  Also,
separate coordinate mappings can be applied to the two grids so that their
resolutions can be controlled separately.

\section{Tests}\label{tests}

Our FD code has been successfully tested by evolving multi-dimensional
shocks.  In order to test the full FD plus PS code, we evolve
equilibrium polytropes.  We choose the domain decomposition of the PS grid
to consist of a filled sphere (a ``ball'') centered on the star
surrounded by several concentric spherical shells.
  Angular basis functions are spherical harmonics. 
For radial basis functions, we use Chebyshev polynomials on the shells 
and an appropriate set of functions~\cite{mm95} on the ball.

\begin{figure}
\begin{center}
\psfig{file=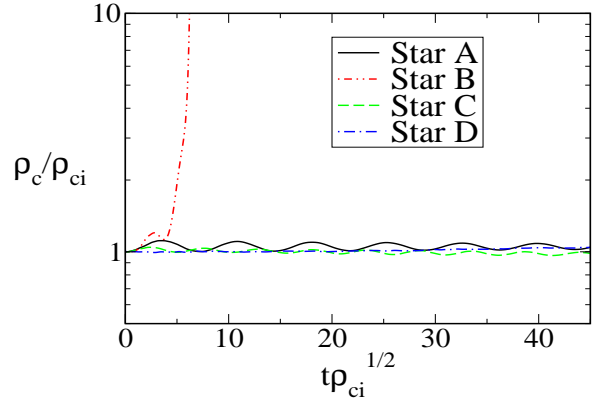,width=2.5in,height=3.5in,angle=270}
\caption{Central density $\rho_c$ as a function of time $t$ 
  for four equilibrium stars, where $\rho_{ci} = \rho_c(t=0)$.}
\label{fig}
\end{center}
\end{figure}

In Figure~\ref{fig}, we show results for four $n=1$ polytropes. 
Star A is a stable
TOV star with central density $\rho_{\rm c}/\rho_{\rm crit}$ = 0.67, star B
is an unstable
TOV star with $\rho_{\rm c}/\rho_{\rm crit}$ = 1.33, star C has the same rest
mass as star A but rotates uniformly with an angular velocity 80\% of the
mass shedding limit, and star C is a hypermassive
rapidly differentially rotating star evolved by other groups~\cite{starC}. 
We choose the FD grid spacing so that 30 points cover a stellar radius. 
(We find that axisymmetric and 3D runs give similar results for all the
models.)  For stars A, B, and C, we choose our PS grids to have
two shells, with the inner one containing the stellar surface.  We use
spherical harmonics up to $L=7$, and we use 7, 9, and 7 radial collocation
points in the ball and the two shells, respectively.  We find that
the error in the PS grid is dominated by the shell containing the stellar
surface, and it decreases quadratically with the grid spacing in
this shell.  We add a 1\% pressure depletion to test stability.  For star D,
we choose the ball with spherical harmonics up to $L=14$ and 18 radial points
to cover the whole star.

We find that our code can accurately evolve equilibrium stars and distinguish
stable from unstable configurations.  We have also
successfully evolved moving stars using the dual coordinate system to track
the star's center of mass.  These tests encourage us to think that our code
might be able to produce accurate simulations of NSNS and BHNS binaries.

\section{Acknowledgments}

This work was supported in part by a grant from the Sherman Fairchild
Foundation, by NSF grants DMS-0553677, PHY-0354631, and NASA grant
NNG05GG51G.

\vfill

\end{document}